\documentclass[12pt]{iopart}

\usepackage{iopams}
\usepackage{graphicx}
\usepackage[latin1]{inputenc}
\usepackage{amssymb}
\begin{document}

\title{Kinetic Monte Carlo modelling of dipole blockade in Rydberg excitation experiment}
\author{Amodsen Chotia, Matthieu Viteau, Thibault Vogt, Daniel Comparat and Pierre Pillet}
\address{Laboratoire Aimé Cotton, CNRS, Univ Paris-Sud, Bât. 505, 91405 Orsay, France}
\eads{\mailto{amodsen.chotia@lac.u-psud.fr}, \mailto{daniel.comparat@lac.u-psud.fr}}
 
\begin{abstract}
We present a method to model the interaction and the dynamics of atoms excited to Rydberg states.
We show a way to solve the optical Bloch equations for laser excitation of the frozen gas  in good agreement with the experiment. A second method, the Kinetic Monte Carlo method gives an exact solution of rate equations. Using a simple N-body integrator (Verlet), we are able to describe dynamical processes in space and time. Unlike more sophisticated methods, the Kinetic Monte Carlo simulation offers the possibility of numerically following the evolution of tens of thousands of atoms within a reasonable computation time. The Kinetic Monte Carlo simulation gives good agreement with dipole-blockade type of experiment. The role of ions and the individual particle effects are investigated. 
\end{abstract}


\section{Introduction}

From the seventies, physics of the Rydberg atoms has been an object of
great interest. Most of the properties of Rydberg atoms are due to the
dimension of the Rydberg orbit, typically in atomic units of the order of
the square of the principal quantum number $n$. Possessing huge electric
dipole moments, large lifetimes..., Rydberg atoms have offered the
opportunity of studying atoms in extreme experimental conditions, for
instance in presence of high electric, magnetic or electromagnetic fields or approximating the 
conditions which corresponds to (low $n$) atoms in the 
neighborhood of a star \cite{Gallagher}. More recently the physics of Rydberg
states of atoms in cold gases has stimulated interest since they are at the frontier of atomic,
molecular, solid-state and plasma physics. In an ensemble of cold Rydberg
atoms many-body phenomena have been observed in a F\"{o}rster configuration
where Rydberg atoms can exchange internal energy through long-range
dipole-dipole interactions \cite{Mourachko,Anderson}. The possibility of
controlling those strong interactions between atoms has been demonstrated by 
using an external controllable electric field \cite{Fioretti}. 

A basic difference between experiments with cold Rydberg atoms and those
with Rydberg atoms at room temperature is that cold Rydberg atoms on the ($\sim $1 $\mu $s) timescale  of the experiments can be approximately considered motionless. For example, in the case of cesium
atoms they move 100 nm which is roughly the size of the atom, for $n\sim 30$. Such a
cold gas is expected not to exhibit any collisions and presents totally different
characteristics than does a thermal gas at room temperature. No collisions does not
mean no interactions, and a frozen Rydberg gas can present novel properties,
close to those of an amorphous solid. The frozen Rydberg gas approximation
leads to considering the ensemble of Rydberg atoms as interacting at large
distances by the van der Waals interaction or the dipole-dipole one. The F%
\"{o}rster configuration leads to a situation very similar to the migration
of excitons \cite{Mourachko,Anderson}. Thus, fascinating perspectives are expected with cold Rydberg atoms.
Controllable long-range interactions are particularly exciting
for quantum information applications especially the so called dipole blockade mechanism of the
excitation due to the strong interactions between Rydberg atoms \cite{Jaksch, Lukin}. The energy of a pair of interacting Rydberg atoms is shifted by dipole-dipole interactions and is not twice the energy of one Rydberg atom. A limitation of the excitation is expected when the dipole-dipole energy shift exceeds the resolution
of the laser excitation. The use of a dipole blockade of the excitation
constitutes an efficient way for the realization of a CNOT quantum gate
\cite{Jaksch, Lukin, Tong, Afrousheh, VogtPRL2006, VogtPRL2007}. The possibility of observing the dipole blockade of laser excitation has been demonstrated for the first time with the van der Waals case \cite{Tong,Afrousheh} and the F\"{o}rster
case \cite{VogtPRL2006}. Modelling the complex behavior of the dipole-dipole interaction in a frozen gas opens interesting ways of understanding the role of each particle by switching on and off the different interactions or effects. The advantage offered by our simulation is the possibility of selectively adding effects/interactions depending on their rates with up to thousands of particles under reproducible conditions within a computational time of a  few minutes. 
After a review of our experimental conditions, we describe 
the different methods we have been using to model the dipole blockade effect observed in \cite{VogtPRL2006,VogtPRL2007}. 
We briefly explain a first method based on the solution of the optical Bloch equations, then we discuss the use of a Kinetic Monte Carlo (KMC) simulation. We then present the results we obtained with the KMC model for different experimental situations. Due to the wide utility of algorithms used, we have presented
in appendix a review of KMC method itself and the algorithm for the motion of the particles.

\section{Experimental setup}

In many experiments with hot or cold Rydberg atoms, the experimental
procedure is the following. The atoms are excited by a short laser pulse ($%
\sim 10$ ns) to a Rydberg state, $nl$ ($l=s,p,d$). Then after a duration of
a few microseconds the Rydberg gas sample is selectively state analyzed by
using a high voltage pulsed electric field with a risetime of the order of
a microsecond. An important difference is observed between experiments
realized at room temperature, using for instance a thermal atomic beam, and
those realized with a cold atomic sample provided by a magneto-optical trap
(MOT). In the case of cold atoms large
fluctuations of the number of Rydberg atoms are generally observed between laser-shots. The
reason is the very narrow linewidth ($\sim 1$ kHz if excited from the ground state, $\sim 5$ MHz if excited from the $6p$ in cesium) of the Rydberg resonance,
compared to the broad bandwidth multimode laser which cavity modes oscillates randomly (multiple cavity modes spread over a few GHz). It leads to uncontrollable frequency shifts of $\sim 500$ MHz. In
an atomic beam, the Doppler
effect can be up to 1 GHz which limits the fluctuations of the
Rydberg population. In the case of the atoms in a MOT, there is no Doppler
effect, which explains the strong fluctuations. In broadband experiments the excitation of an ensemble of Rydberg atoms
interacting altogether corresponds to the excitation of a band of energy levels,
which can be excited by a short, thus broadband, laser pulse. The width of the band versus
the Rydberg atomic density has been investigated by microwave spectroscopy
\cite{Afrousheh} and laser spectroscopy \cite{MudrichPRL05}. Using monomode
lasers for the excitation of cold atoms is a way to avoid the fluctuations
of the Rydberg population from shot to shot.

Another important difference is expected between a broadband excitation and a
high-resolution one. With narrow band, low power excitation, only a small part of the band of levels can be excited, leading to
the limitation of the excitation and corresponding to a van der Waals \cite{Tong,Singer} 
or dipole \cite{VogtPRL2007} blockade. The first excited Rydberg atoms shifts the
resonance of the non-excited neighbors and prevent their excitation in a
narrow-bandwidth laser excitation.

\begin{figure}[ht!]
	\centering
\resizebox{0.6\textwidth}{!}{
		\rotatebox{-90}{\includegraphics*[20mm,37mm][110mm,198mm]{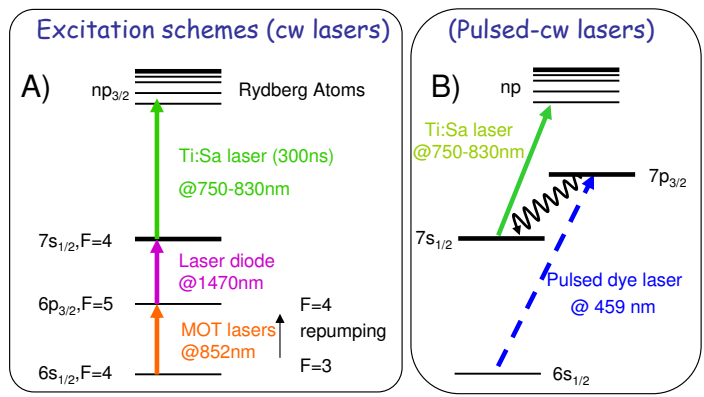}}}
\caption{(A) Three step excitation scheme for Cs Rydberg atoms. (B) Combined pulsed and
cw-excitation.}
	\label{fig:excscheme}
\end{figure}

The details of our experimental setup have been described in several papers 
\cite{Mourachko,VogtPRL2006,Fioretti,MudrichPRL05}. The Rydberg atoms are
excited in a cloud of up to $5\times 10^{7}$ Cs atoms (temperature 200 $\mu $%
K, characteristic radius $\sim $\ 300 $\mu $m, peak density $1.2\times
10^{11}$\ cm$^{-3}$) produced in a standard vapor-loaded MOT at residual gas
pressure of $3\times 10^{-10}$\thinspace mbar \cite{Mourachko,Fioretti}. At
the trap position, a static electric field and a pulsed high voltage field
can be applied by means of a pair of electric field grids spaced by
15.7\thinspace mm. We consider an ensemble of cold cesium atoms excited in
the Rydberg state, $np_{1/2}$\ or $np_{3/2}$. Three cw lasers provide
a high resolution multistep scheme of excitation, as depicted in figure~\ref{fig:excscheme} A).

 The first step of the excitation, $6s,F=4\rightarrow
6p_{3/2},F=5$, is provided by the trapping lasers (wavelength: $\lambda
_{1}=852$ nm) or a diode laser to avoid the excitation of hot atoms.\ The
density of excited, $6p_{3/2}$, atoms can be modified for instance by
switching off the\ repumping lasers before the excitation sequence.\ The
second step, $6p_{3/2},F=5\rightarrow 7s,F=4$, is provided \ by an infrared
diode laser in an extended cavity device (wavelength: $\lambda _{2}=1.47$ $%
\mu $m, bandwidth: 100 kHz and available power: 20 mW). The average experimental intensity is $\sim 3$ mW/cm$^{2}$, twice the saturation one. The last
step of the excitation, $7s,F=4\rightarrow np_{1/2,3/2}$ (with $n=25-300$),
is provided by a Titanium:Sapphire (Ti:Sa) laser. The wavelength $\lambda _{3}$
ranges from $770$ to $800$ nm, the bandwidth is $1$ MHz, and the available
power is $400\,$mW. 
The Ti:Sa laser is switched on for a time, $\tau =0.3\,\mu $s, by means of an acousto-optic
modulator at an $80$Hz repetition rate. The beams of the infrared diode laser and of \ the Ti:Sa laser
cross with an angle of 67.5\ degrees and are focused into the atomic cloud
with waists of 105 and 75 $\mu $m, respectively. Their polarizations are
both linear and parallel to the direction of the applied electric field,
leading to the excitation of the magnetic sublevel, $np_{1/2}$ or $np_{3/2}$ 
$\left\vert m\right\vert =1/2$. The spectral resolution, $\Delta \nu _{L}$,
of the excitation is $5-6$ MHz, limited by the lifetime,
56.5 $ns$, of the $7s$\ state and by the duration, i.e by the spectral width of
the Ti:Sa laser pulse.\ The magnetic quadrupole field of the MOT is not
switched off during the Rydberg excitation phase, but it contributes less
than 1 MHz to the observed linewidths. Just after the Ti:Sa laser pulse (between
$0$ and $1µs$) the Rydberg atoms are selectively
ionized  by applying a pulsed high-voltage field with a rise time of 700 ns and detected on a Micro Channel Plate (MCP)
detector. The experimental procedure is
based on spectroscopy of Stark $np$ states for
different atomic densities, and for different Ti:Sa laser intensities.

\section{Dipole blockade model}

Different approaches have been followed  to study the problem of the excitation to a Rydberg state in presence of already excited atoms \cite{Robicheaux2005, Ates2007, Amthor2007}. We discuss hereafter some hypotheses and simplifications we made to model the blockade effect. A first simplification is made by considering the excitation to $np$ states only. The dipole-dipole interactions are calculated for first and second orders between the $np$ and all the neighboring states in the Stark diagram. When looking at a specific atom to be excited, the shift in energy is due to already  excited neighboring atoms but not ground state atoms. As shown in the next section, the sum of each individual atom's contribution can be studied and the main effect is due to the nearest neighbor.
  
If a static electric field is present, either external or due to ions, Rydberg states are mixed, creating a permanent dipole moment for the Rydberg atoms.  
For instance in Cesium in the presence of an electric field $\vec{F}$, the $np$ state is mainly mixed with the $(n-1)d$ state. We denote by $µ_{pd}=\left\langle np_j,m_j=1/2\left|q_e z\right|(n-1)d_{j+1},m_{j+1}=1/2\right\rangle$ the transition dipole moment of an atom in state $np$ ($j,m_j=1/2$) toward $(n-1)d$ ($j+1,m_{j+1}=1/2$).
We introduce the scale parameter $\theta $ characterizing the dipole coupling for each level $np$
defined by $\tan \theta =\frac{\mu_{pd}.F}{h\Delta _{pd}/2},
$ where $h\Delta _{pd}$ is the zero field energy difference between the $(n-1)d$ and $np$ levels. Energies and dipoles are obtained following \cite{Zimmerman}, and are calculated only for  $|m_j|=1/2$ states.
The dipole of an atom in a $np$ state aligned along the local electric field $(\vec{F})$ is given by (here z is the coordinate along the vector defined by $\vec{F}$):
\begin{equation}
	\mu_{pd}(F)=<np(F)|q_e z|np(F)> = \mu_{pd}\sin\theta
\end{equation}

Where the basis ($|np(\vec{F})>$,$|(n-1)d(\vec{F})>$) are the eigenstates given by the diagonalization of the Hamiltonian matrix

\begin{equation}
\left( \begin{array}{ccc}
E_p & -\mu_{pd}F\\
-\mu_{pd}F & E_d  
\end{array} \right)
\end{equation}
where $E_p$ and $E_d$ are the energies of states $np$ and $nd$ in absence of an electric field. The resulting shift in energy for $np$ is $h\Delta_p(\vec{F})=\frac{h\Delta_{pd}}{2}(1-\sqrt{1+\tan^2(\theta)})$, where $h\Delta_{pd}=E_p-E_d$.
 
One can calculate the dipole-dipole interaction term $V_{ij}$ between two atoms labelled $i$ and $j$ separated by $\vec{R_{ij}}=R_{ij}\vec{n_{ij}}$.
The first order dipole-dipole interaction is :
\begin{eqnarray}
 V_{ij}^{dip}=\frac{\overrightarrow{\mu}(\vec{F_{i}}).\overrightarrow{\mu}(\vec{F_{j}})-3\left( 
\overrightarrow{\mu}(\vec{F_{i}}).\overrightarrow{n}_{ij}\right) \left( 
\overrightarrow{\mu}(\vec{F_{j}}).\overrightarrow{n}_{ij}\right) }{4 \pi \epsilon_{0}
R_{ij}^{3}}
\label{eq:dipdip}
\end{eqnarray}
where $h\Delta_{pk}$ is the energy difference between states $np$ and $n'k$, and $\vec{µ}(\vec{F_i})=µ_{pd}(\vec{F_i}).\frac{\vec{F_i}}{\left\|\vec{F_i}\right\|}$ is the classical permanent electric dipole of atom $i$ which is aligned along the local electric field $\vec{F_i}$.
In the absence of an electric field, $\theta=0$ and there is no permanent dipole moment and only the second order, so called van der Waals interaction is non zero.
The potential energy of a $np$ Rydberg atom is the sum of the energy of the state without electric field $E_p$ plus the shift of the state in the local electric field $\Delta_p(\vec{F_{i}})$ plus the sum of the dipole-dipole interactions with all the atoms. For the atom $i$ we then note
\begin{eqnarray}
E_{pot}[i]=E_p+ h\Delta_p(\vec{F_i})+\sum_{j\neq i}{V_{ij}}\nonumber \\
E_{pot}[i]=E_p+h\frac{\Delta_{pd}(i)}{2}(1-\sqrt{1+\tan^2(\theta_i)})+\sum_{j\neq i}{V_{ij}}
\label{eq:ep}
\end{eqnarray}

An important part of the computation time is used to calculate the local electric fields and the potentials.

\subsection{Reduced density matrix, mean field simulation}

The details of this work can be found in \cite{These_Thibault}. We just briefly review the main results.
The process describing the three excitation steps (see figure \ref{fig:excscheme} A)) for one atom can be described using the optical Bloch equations. 
\begin{equation}
	\frac{d\rho}{dt}=-\frac{i}{\hbar}[H,\rho]-\frac{1}{2}(\rho\Gamma+\Gamma\rho)+\gamma
	\label{eq:bloch}
\end{equation}	
Where the time evolution of the density matrix $\rho$ is decomposed into three terms.
The first term contains an Hamiltonian H being the sum of the individual potential energies, and the interaction of an atom with all the others $\sum{V_{ij}}$ in presence of electric fields $H=\sum_i E_{pot}[i]$. The second term accounts for the relaxation of the populations and coherences where $\Gamma$ gives the lifetime of the considered state. The third term $\gamma$ takes into account the radiative relaxation to state $l$ with energy $E_l$ from states k with energy $E_k>E_l$ due to spontaneous emission.
Taking the trace over all the atoms except the one labeled $i$ in the optical Bloch equations, gives the evolution of the density matrix for the particle $i$.  The interaction term or shift in energy for the atom $i$ due to the interaction with its neighboring atoms is $\sum_{j\neq i}{Tr_{j}[V_{ij},\rho_{i,j}]}$, with $V_{ij}$ the dipole-dipole interaction and $\rho_{i,j}$ the two-body density matrix for atoms i and j. The coupling with all the other atoms becomes a mean field term proportional to the atom density. A similar treatment has been performed by \cite{Ates2006}.

 As correlations appear during the excitation, the state of the system does not remain a product state. However the probability of excitation of a ground state atom into a Rydberg state being on the order of a few percent, and as long as the product of the individual density matrices is small, we can use the Hartree-Fock approximation. In this approximation the two-body density matrix $\rho_{i,j}$ can be developed as the product of single atom density matrices. We start with one atom $i$ in ground state and atom $j$ in the excited state, denoted (ge). After excitation, the system ends in a double excitation noted (ee). The Hartree-Fock approximation allows us to write, $\rho_{(i,j)ge,ee}=\rho_{i_{ge}}\times\rho_{j_{ee}}$. 

\begin{figure}[h!]
	\centering
	\resizebox{0.9\textwidth}{!}{
		\rotatebox{-90}{\includegraphics*[22mm,35mm][110mm,260mm]{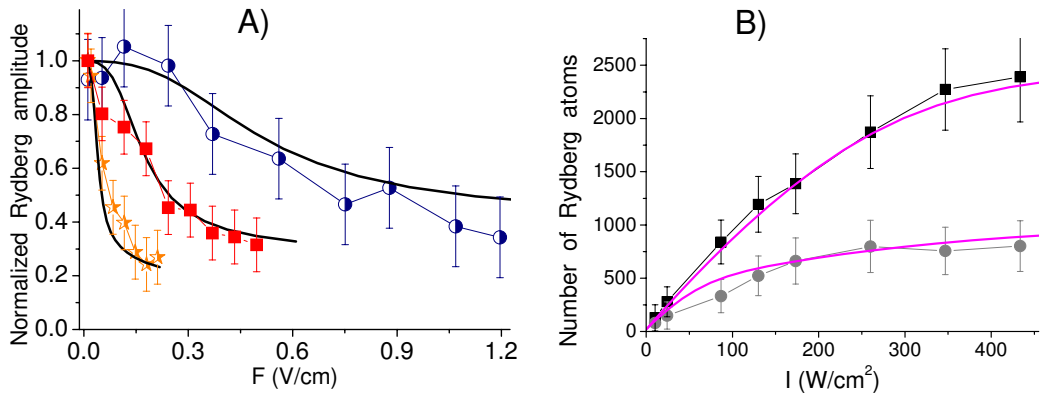}}}
		\caption{(A) Probability in our sample, as a function of the
electric field $\vec{F}$, of an atom to be excited in Rydberg state compared to the
isolated atom probability excitation. $n$ is equal to 60 (circles), 70
(squares)and 85 (stars). The Ti:Sa laser intensity is given by 
$(n/85)^3 \times 560$ W/cm$^2$. Symbols represent experimental data and solid lines represent the reduced density
matrix model.(B) Number of Rydberg atoms excited
versus the Ti:Sa laser intensity, in the case of the $70p_{3/2}$ state, for
$7s$-atom density $D\sim 4 \pm 2 \times 10^9 cm^{-3}$ and in the
presence of two different electric fields, 0 V/cm (squares) and 0.25
V/cm (circles). Solid lines show the reduced density
matrix model taking into account the van der Waals blockade at
zero field and the dipole blockade in the presence of the electric field.}
	\label{fig:meanfield}
\end{figure}

 Thus the interaction term can be written as 
\begin{equation}
	\delta_{dd}(i)=\sum_{j\neq i}{Tr_{j}[V_{ij},\rho_{i,j}]}=(\sum_{j \neq i}V_{ij}\rho_{j_{ee}})(\rho_{i_{ge}}|g_i><e_i|-\rho_{i_{eg}}|e_i><g_i|)
\label{eq:interaction}
\end{equation}
which is simply a shift of the Rydberg level for the atom $i$. As an illustrative example, we look at a weak interaction with $\tan^2(\theta)$=0.05, and with a $70p_{3/2}$ state. The dipole blockade effect induces a shift of 6MHz (exactly our excitation linewidth) which prevents the excitation of two atoms at a distance of 5$µ$m. At a density of $10^{11}$cm$^{-3}$ a sphere of radius 5$µ$m contains 50 atoms. This means that only one excitation could be present for the 50 atoms, and the probability to excite the atom $j$ would be uniform within this sphere. 
The population in the excited state $\rho_{j_{ee}}$ is then  replaced by a mean value $\overline{\rho_{ee}}$. Due to the inhomogeneity of the atomic density and laser intensity a local  $\overline{\rho_{ee}}(\vec{r})$ is considered at different positions $\vec{r}$ over the whole atomic cloud.
A naive (mean field) estimation for $\rho_{j_{ee}}$ could lead to wrong estimations. Indeed, a mean field interaction for an atom at the center of the cold atomic could naively be written as the integral of the interaction term $V_{ij}$ over all the possible directions $\Theta$ (the angle between the internuclear axis and the direction of the dipole $i$): $\int^\pi _0{V_{ij} sin\Theta d\Theta}$ which is equal to zero, but is not the real value. In order to overcome this problem, a better way to evaluate the local interaction potential is to consider separately the nearest neighbor Rydberg atom from the other atoms. The shift in energy $\delta_{dd}(i)$ is decomposed into a sum over all the atoms $j$ treated as a continuous distribution out of a sphere containing only one excited atom (the nearest neighbor of $i$) in its center, plus the contribution of the nearest neighbor Rydberg atom. The result from this calculation is that the local field contribution associated to the nearest neighbor is dominant over the mean field contribution if an electric field is present. The nearest neighbor Rydberg atom contribution is considered at the most probable distance given by the mean of the Erlang distribution \footnote{The probability to find a $k^{th}$ nearest neighbor at a distance $r$ of a an atom is given by  $4\pi r^2*f(k,r)$, with $f(k,r)=\frac{3}{4\pi k!}\frac{{(r^3)^{k-1}}}{(Rd^3)^{k}}e^{-(\frac{r}{Rd})^3}$ and $Rd=(\frac{4\pi \rho_0}{3})^{-\frac{1}{3}}$.} \cite{Torquato} from the atom $i$. The shift in energy relies on the local density (gaussian distributed) $\rho_0(\vec{r})$ of the atoms in ground state. We finally find that the shift for the atom $i$ is given by  
\begin{equation}
	\delta_{dd}(i)\propto\overline{\rho_{ee}}(\overrightarrow{r})\rho_0(\vec{r})
	\label{eq:shift}
\end{equation} 
 We then solve equation (\ref{eq:bloch}) for an atom $i$ using the result from equation (\ref{eq:shift}). The result given in figure \ref{fig:meanfield} reproduces well the experiment. We take into account multiphotonic excitations as well as the finite coherence time of the lasers in the model through a temporal phase variation of the electric field of the lasers in the three step excitation represented in figure \ref{fig:excscheme} A). Two results are given in figure \ref{fig:meanfield}, where the reduced density matrix approach is plotted versus the experimental data for different electric fields A) and different intensities B).

\subsection{Kinetic Monte-Carlo (KMC) simulations}
\label{KMC_algo}

 The previous approach based on the reduced density matrix has some limitations. Despite correctly describing the excitation it was not possible to look for  the dynamics of the system, the orientation of the dipoles in a local electric field or the individual interactions between atoms instead of a mean field term or ionization. For these reasons we developed a KMC simulation. All the above mentioned limitations can then be overcome. However in KMC simulations the excitation has to be based on the solution of rate equations. A more detailed description of the KMC algorithm, and more generally of possible numerical solution of any kind of master or rate equations, is given in \ref{appendixA}. 
Briefly if a system is driven by a master equation
\begin{equation}
    \frac{dP_k}{dt} =  \sum_{l=1}^N \Gamma_{k l} P_l - \sum_{l=1}^N  \Gamma_{l k}  P_k 
    \label{mastereq}
\end{equation}
describing the time evolution of the probability $P_k$ of a system to occupy each one of a discrete set of states numbered by $k$.  Each process occurs at a certain average rate $\Gamma_{l k}$, which may either
be constant in time, or dependent on how the system has evolved up to that time.

The KMC algorithm is then  the iteration of the following steps.
\begin{itemize}
\item Initializing the system to its given state called $k$ at the actual time $t$.
	\item Creating the new rate list $ \Gamma_{l k} $ for the system, $l=1,\ldots,N$.
	\item  Choosing a unit-interval uniform random number generator \footnote{In our case we use the free 
 implementations by GSL (GNU Scientific Library) of the Mersenne twister unit-interval uniform random number generator  of Matsumoto and
Nishimura.} $r$: $0<r\leq 1$ and  calculating the first reaction rate time $t'$ by solving
	$ \int_{t}^{t'} \sum_{l=1}^N \Gamma_{l k} (\tau) d \tau =  - \ln r$.
	\item Choosing a unit-interval uniform random number generator  $r'$: $0<r'\leq 1$ and searching for the integer $l$ for which $R_{l - 1} < r' R_N \leq  R_l$ where
 $R_j = \sum_{i=1,j} \Gamma_{i k} (t')$ and $R_0=0$. This can be done efficiently using a binary search algorithm.
	\item Setting the system to state $l$ and modifying the time to $t'$. Then go back to the fist step.
	
\end{itemize}

One fundamental result is that the KMC method makes exact numerical calculations and cannot be distinguished from an exact molecular dynamics simulation, but is orders of magnitude faster. It is therefore indistinguishable from
the behavior of the real system (if evolving through a master equation), reproducing for instance all possible data in an experiment including its statistical noise. 

\subsection{KMC simulation of dipole-dipole interaction}
We consider a cloud with a gaussian spatial density. Initially the thousands of atoms are in their ground state with a maxwellian distribution for the velocity $\sigma_v=\sqrt{k_b T/m_{at}}$, where $T,k_b,m_{at}$ are respectively the temperature of the frozen gas in the MOT, the Boltzmann constant and the mass of the atomic species under consideration.
In such a system, considering coherent excitations would
lead to solving the Schr\"{o}dinger equation with $\approx 2^{100}$ states, which is obviously beyond our capability. Nevertheless, it is possible to obtain good agreement for the shift in energy and the dynamics of the system with a reasonable computation time using a simplified numerical treatment. We consider a two level system for each atom so they can be either in their ground state $(6s_{1/2})$ or laser excited to a given $(np_{3/2})$ Rydberg state. Initially, the electric dipoles $\vec{µ}$ of the atoms are aligned along the  direction of polarization $z$ of the exciting laser and the applied DC electric field, which is the quantization axis. During the evolution (ionization especially) the electric dipoles will be aligned along the local electric field. The shift $\delta_{dd}(i)$  of the Rydberg state of atom $i$ is then $\sum_{j\neq i} V_{ij}$ where $V_{ij}$ is given by equation (\ref{eq:dipdip}).
We start from the sets of Bloch equations for our two level system \cite{BluchsAuzinsh}:
\begin{eqnarray}
\frac{d\rho_{ee}}{dt}=-\Gamma_{spont}\rho_{ee}-\frac{i}{2}\Omega(\rho_{ge}-\rho_{eg})\\
\frac{d\rho_{ge}}{dt}=-(\frac{\Gamma_{laser}+\Gamma_{spont}}{2}-i(\delta_{laser}+\delta_{dd}))\rho_{ge}+\frac{i}{2}(\rho_{gg}-\rho_{ee})\Omega\\
\frac{d\rho_{gg}}{dt}=\Gamma_{spont}\rho_{ee}+\frac{i}{2}\Omega(\rho_{ge}-\rho_{eg})
\end{eqnarray}

where $\rho_{gg}$ and $\rho_{ee}$ are the populations of ground state and excited state atoms, $\Gamma_{spont}$ is the spontaneous decay rate, $\Gamma_{laser}$ is the FWMH of the exciting laser($\Gamma_{spont}<<\Gamma_{laser}$), $\Omega$ is the local Rabi frequency of the transition and $\delta_{laser}$ is the detuning from the isolated and field free atom resonance. In order to end up with rate equations for each atom, we neglect the coherences $\dot{\rho_{ge}}=\dot{\rho_{eg}}=0$. This is obviously the biggest assumption made here. Coherences are in fact tractable with a Monte Carlo method as described in \cite{Minami}. We finally get pure rate equations for $\rho_{ee}$ and $\rho_{gg}$:
\begin{eqnarray}
	\frac{d\rho_{ee}}{dt}=-(\Gamma_{spont}+\Gamma_{exc})\rho_{ee}+\Gamma_{exc}\rho_{gg}\\
	\frac{d\rho_{gg}}{dt}=-\Gamma_{exc}\rho_{gg}+({\Gamma_{spont}+\Gamma_{exc}})\rho_{ee}\\
	\Gamma_{exc}=\cos^2(\theta_i/2)\frac{\Omega^2\Gamma_{laser}}{{\Gamma_{laser}}^2+4(\delta_{laser}+\delta_{dd})^2}	
\end{eqnarray}
where the rate $\Gamma_{exc}$ of excitation of atom $i$ depends on the projection of $|np(\vec{F_i})>$ onto $|np>$, which is $\cos^2(\theta_i/2)$, times the stimulated laser rate. 
The deexcitation rate is similar to the excitation rate plus the spontaneous decay of the $np$ state. 
The great advantage offered by this 'kinetic' method, is the simplicity of adding phenomena or particles with evolution based on rates.   
Contrary to the previous model the potential energy of an atom is the sum of the energies due to the Stark shift and the interactions with all the other atoms.\\ 
Numerical methods such as the ordinary Runge--Kutta methods are not ideal for integrating Hamiltonian systems because they do not conserve energy. On the contrary symplectic integrators such as the Verlet integrator does conserve energy. Mechanical effects are then treated via classical movements of the atoms, and dipolar forces between atoms are taken into account using the (leapfrog) Verlet-Störmer-Delambre algorithm (see \ref{appendixB}). 
The computation is realized as follows. 
Ion formation is possible and creates an electric field felt by other atoms, thus the direction and strength of the dipoles can vary depending on local electric fields. Two main mechanisms for ionization exist. The first one is due to laser ionization or blackbody ionization and has a rate of ionization proportional to the atomic density. The second one happens if two Rydberg atoms move toward each other and reach a smaller internuclear distance than $~4n^2a_0$ \cite{Robicheaux2005ionization}. In the latter case one Rydberg atom is ionized, the second atom falls to a lower state. Due to energy conservation, its binding energy is at least twice as large after the ionization, and as the final atomic states are often different from $s$ or $d$, it does not interact with $np$ atoms. Consequently we assume for simplicity in the simulation that the atom state is changed to a non interacting state. 

Due to its time and spatial resolution the KMC simulation takes into account all the dipolar interactions developed during the excitation and gives access to individual atoms. A dynamical evolution of the system is made except if a collision between two atoms in $np$ state is detected or if a reaction is detected. 
The time evolution of the simulation is incremented either with the KMC timestep or with a small fraction of the collisional timestep. After each change in the position or change of any particle state, the fields and potentials are recalculated over all the atoms. Then operations are repeated until the end of the excitation.

\section{Results}

\subsection{Field induced dipole blockade}

As described in the experiment reported in \cite{VogtPRL2007} the dipole moment of $np$ states increases with the strength of the coupling field and reaches a maximum when $\tan \theta$ is equal to 1.
Indeed, as the strength of the field increases the blockade radius increases and the number of excited atoms in a given volume gets smaller. It is worth noting that the dipole blockade condition in a 2-level approach is $\sum_{j\neq i}V_{ij}>h\delta_{laser}$.

\begin{figure}[ht!]
	\centering
	\resizebox{0.6\textwidth}{!}{
		\rotatebox{-90}{\includegraphics*[30mm,40mm][133mm,186mm]{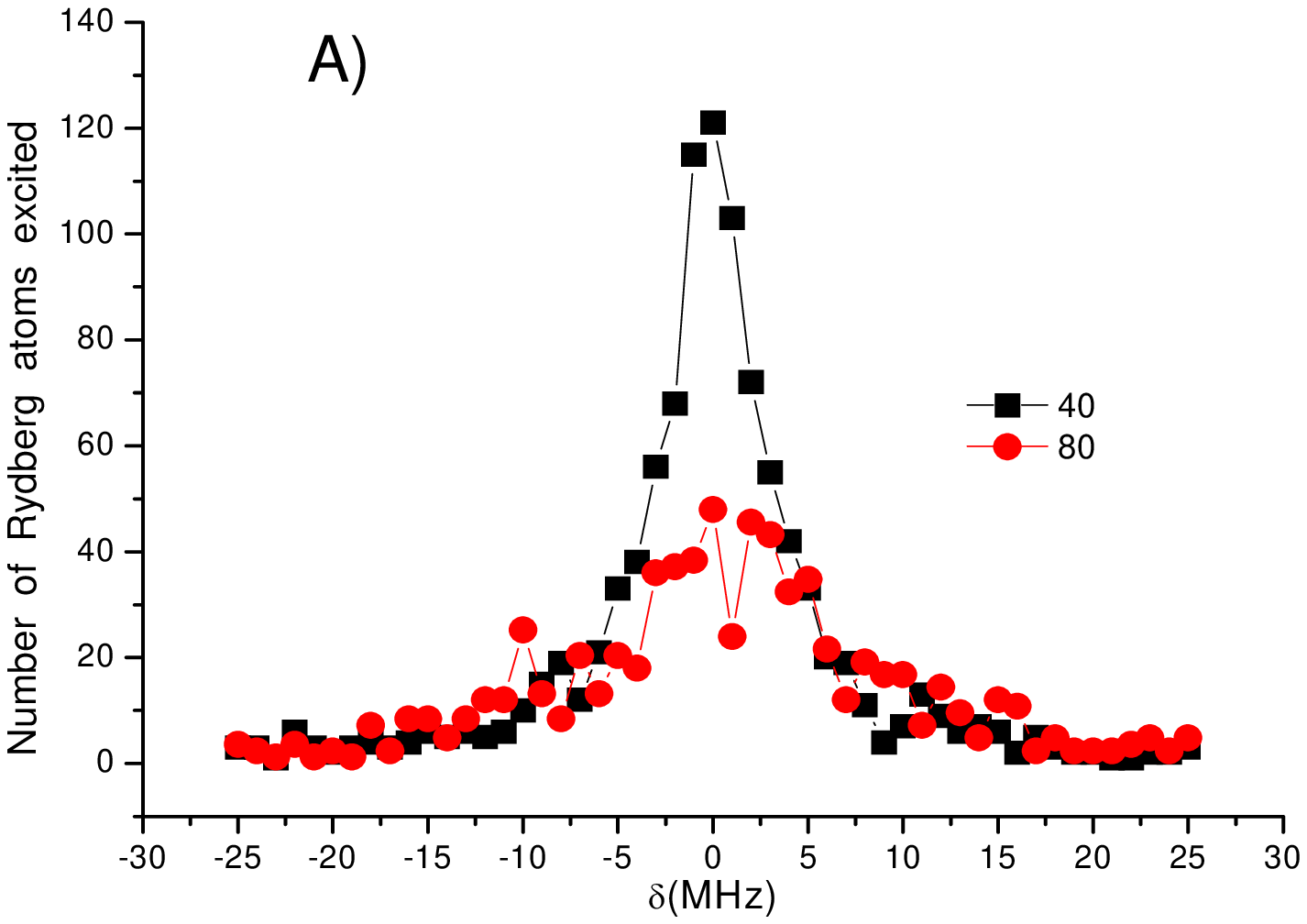}}}
		\resizebox{0.6\textwidth}{!}{
		\rotatebox{-90}{\includegraphics*[23mm,37mm][145mm,201mm]{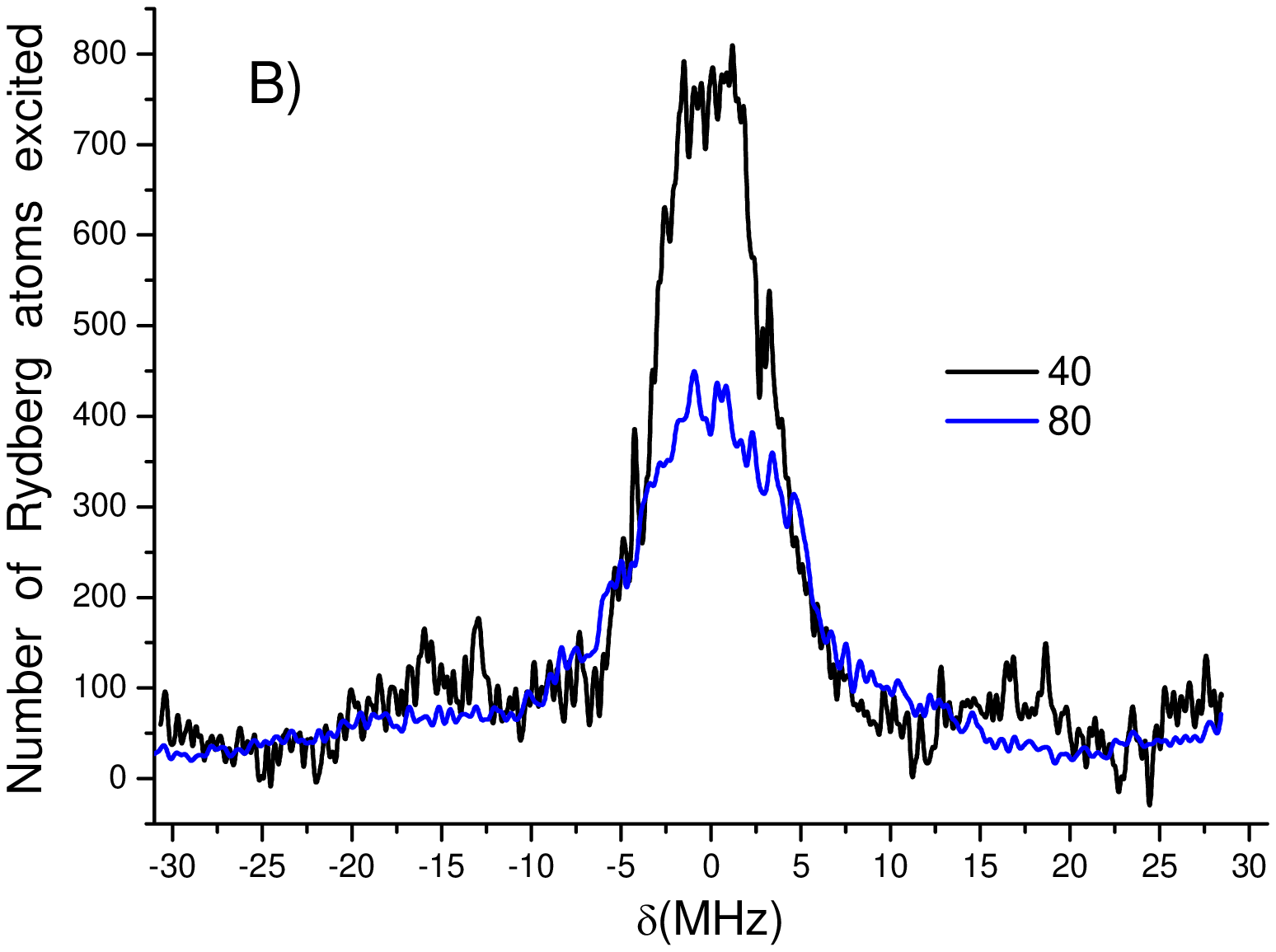}}}
		\caption{(A) Number of Rydberg atoms excited versus the detuning of the excitation laser for n=40 and 80. Monte Carlo simulations with $~2000$ atoms at a density $D=2*10^{10}cm^{-3}$ and an external electric field of $50mV/cm$. This result is taken after a laser excitation time of 300ns, I=0.7*Isat, where Isat is the saturation intensity of the transition $7s\rightarrow np$. The mean number of ions formed during the simulation varies from 0.5 for n=40 to 1.3 for n=80. (B) Experimental data for n=40 (P=20mW) and 80 (P=100mW) at $D=2*10^{10}cm^{-3}$, excitation time $300ns$. The mean number of ions formed per shot is less than 2.}
	\label{fig:champ}
\end{figure}
 
Figure \ref{fig:champ} A) represents the result of a Monte Carlo simulation for different $np$ states, where the number of Rydberg atoms present at the end of the excitation is given as a function of the detuning. The parameters are close to those described in \cite{VogtPRL2007}. Figure \ref{fig:champ} B) shows the results from the experiment. 

\subsection{Effect of the ions}

\begin{figure}[ht!]
	\centering
	\resizebox{0.6\textwidth}{!}{
\rotatebox{-90}{\includegraphics*[19mm,48mm][146mm,220mm]{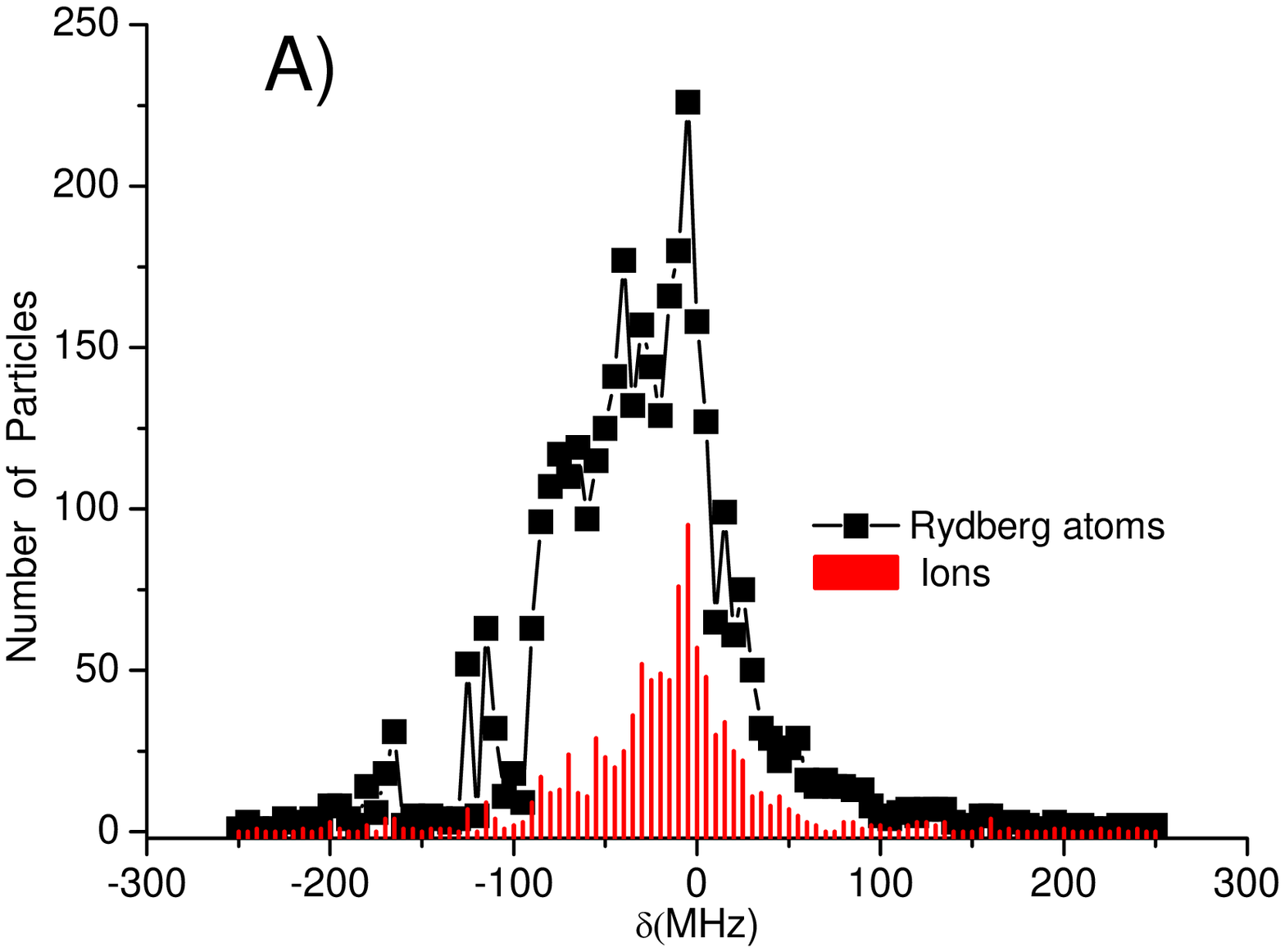}}}
\resizebox{0.6\textwidth}{!}{
		\rotatebox{-90}{\includegraphics*[14mm,24mm][191mm,275mm]{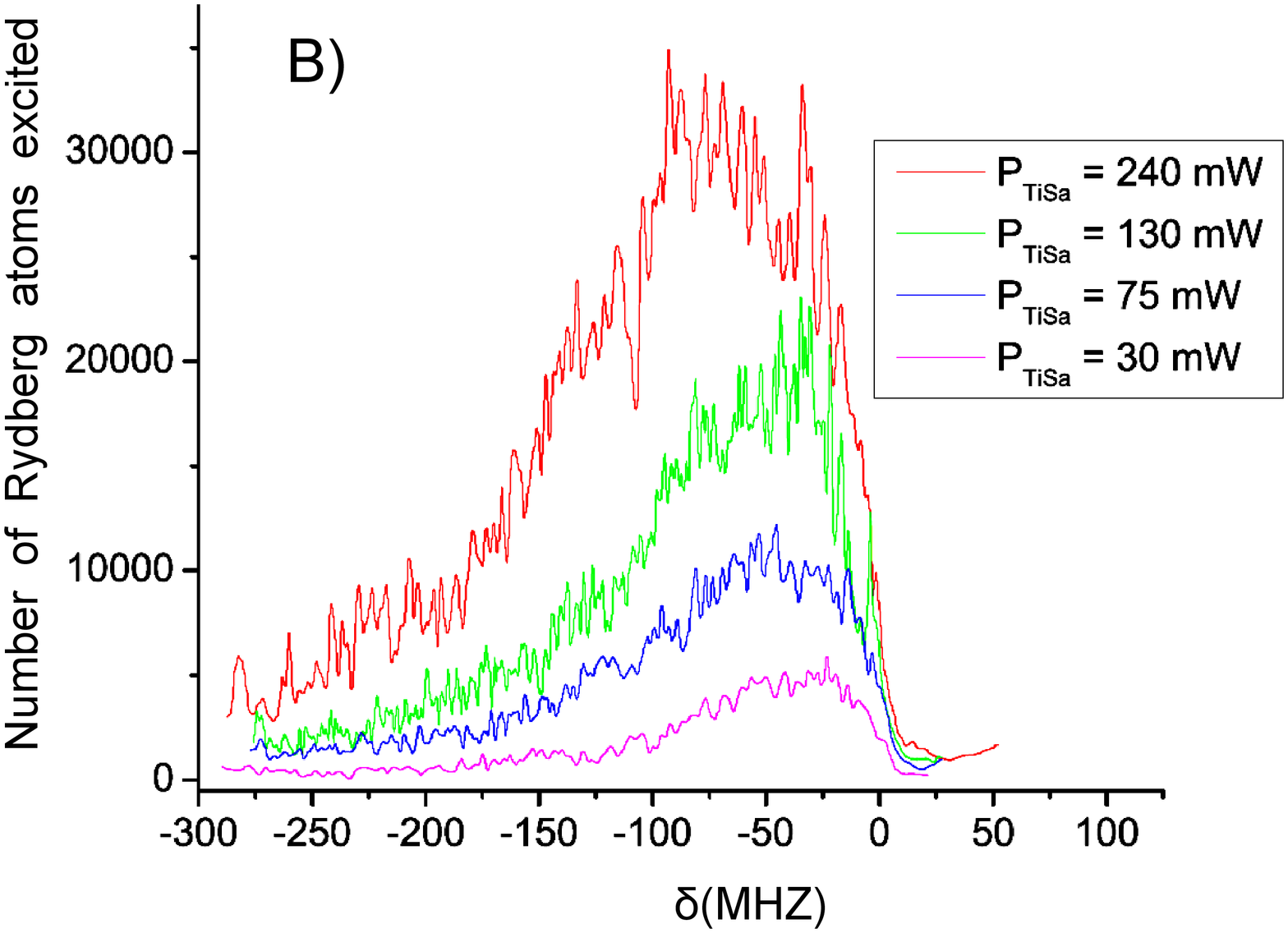}}}
\caption{A) Monte Carlo simulation. Number of Rydberg atoms and ions versus the laser detuning. For $n=81$, excitation time $20µs$, $2000$ atoms, $D=2*10^{10}cm^{-3}$, I=10*Isat. B) Experimental result, see figure \ref{fig:excscheme} B) and \cite{MudrichPRL05} for details on the excitation scheme. Number of Rydberg atoms versus the laser detuning in a combined pulsed ($7ns$,$15µJ$) and cw-excitation (Ti:Sa) at different powers for the excitation of 37p1/2. Excitation time (Ti:sa)= $400ns$.}
	\label{fig:longexc}
\end{figure}

At a distance of $10µm$ the electric field due to an ion is 150mV/cm which represents a shift of 150MHz for an atom in state $70p$. The contribution to the blockade of the excitation of such an ion in the sample during the excitation is so important that it completely hides the observation of the dipole blockade effect. In the experiment the ions present before and during the excitation can be discriminated from the ionized Rydberg atom in the time of flight signal. In the simulation we simply monitor the number of ions at the end of the the excitation time. This ionic effect is shown in figure \ref{fig:longexc}.
In figure \ref{fig:longexc} A) we increase the laser intensity to ten times the saturation intensity and we apply the excitation laser for $20µs$ in absence of external electric field. One can see that the number of excited atoms is important but also that the resonance is broadened to the low frequency side of the atomic resonance. This result is similar to the one obtained in the experiment \cite{Singer}. Ions are formed due to the high laser power but also by collisions during the interaction time due to the long range attractive dipole force between atoms. The number of ions produced during the excitation is so important that it allows for the excitation of Rydberg atoms on a 100MHz range, red detuned from the center of the line. This is due to the fact that $np$ states can only be shifted to the red of the resonance when no external electric field is present. In figure \ref{fig:longexc} B) is shown an experimental curve obtained in a combined pulsed and cw-excitation (see figure \ref{fig:excscheme} B)). In this case the main source of broadening is the inhomogeneous electric fields due to the ions formed by the pulsed laser.

\subsection{Role of the nearest neighbor}
 
 As the dipole-dipole interaction term $V_{ij}$ strongly depends on the distance between two atoms, the energy shift due to the nearest neighbor Rydberg atom has to be distinguished from the mean field shift due to all the other atoms in a Rydberg state. In some cases the contribution of the nearest neighbor can be dominant. 
  
\begin{figure}[ht!]
	\centering
	\resizebox{1\textwidth}{!}{
		\rotatebox{-90}{\includegraphics*[11mm,3mm][119mm,292mm]{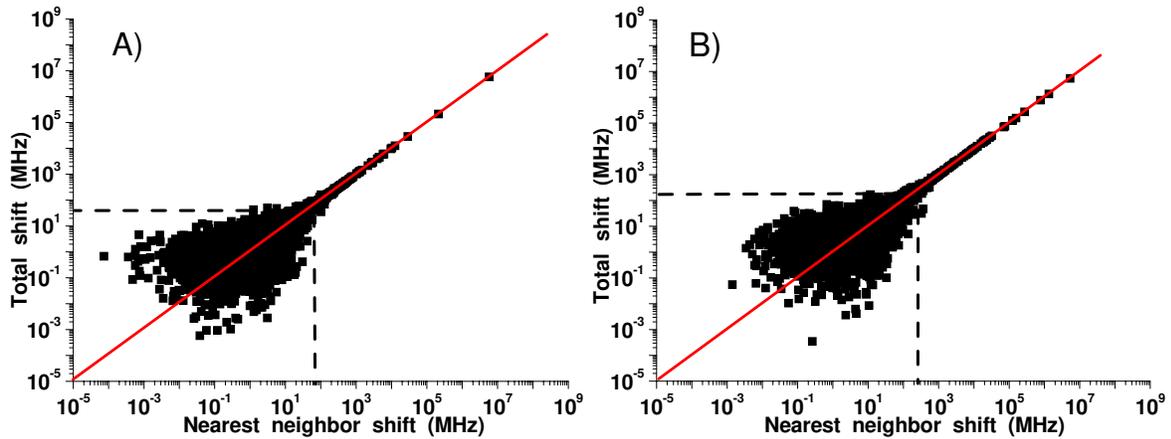}}}
		\caption{Shift of the level of a randomly chosen ground state atom due to the nearest neighboring Rydberg atom compared to the total shift due to all the Rydberg atoms in the Monte Carlo simulation. Simulations are realized for 2000 atoms in a gaussian volume, for an excitation time of 300ns in $n=70$, with a laser intensity of 0.7 Isat. A) D=$5\times10^9 cm^{-3}$, B) D=$5\times10^{10} cm^{-3}$. Along the red line the contribution of the nearest neighbor to the total blockade effect is equal to one. The dashed lines intersection gives roughly a value where the energy shift is only due to the nearest neighbor Rydberg atom.
		}
	\label{fig:nnryd}
\end{figure}
 
 In figure \ref{fig:nnryd} is represented for two different densities, but with no ions to avoid extra effects, the nearest neighbor Rydberg atom shift versus the sum of the shifts of all the Rydberg atoms in the sample. Above the solid line the contribution of the nearest neighbor Rydberg atom to the blockade effect is dominant. The nearest neighbor Rydberg atom shift is twice more important than the shift from all the others Rydberg atoms for $66\%$ of the ground state atoms at a density of D=$5\times10^9 cm^{-3}$ (figure \ref{fig:nnryd} A)), whereas at D=$5\times10^{10} cm^{-3}$ (figure \ref{fig:nnryd} B)) it is $73\%$. This confirm the fact that the energy shift is dominated by the effect of the nearest neighbor. This result has been used to derive the local mean field approximation in the single atom density matrix model previously described.

\section{Conclusion}

In this paper we have presented different methods used to model the dipole blockade effect in a manner as close as possible to an experimental situation. We have first shown that a nearest neighbor mean field analytical approach, based on the solution of the Bloch equations for the partial density matrix equations of one atom of interest gives results in good agreement with the experiment if no ions are present. 
Second the Kinetic Monte Carlo simulation based on rate equations is able to introduce all the electric dipole interactions. Furthermore, we have included N-body spatial dynamics using the Verlet integrator, but the overall code remains very simple. This allows us to look at the important role of the ions formed during the excitation. If ions are present during the excitation they may mimic the dipole blockade effect and lead in extreme cases to a broadening of the lines. We have shown that the two models reproduce well our experiments. It is observed that the number of excited atoms for a given ground state density decreases with the principal quantum number and with the intensity of the electric field which creates the permanent dipole. The effects associated with the experiment, such as the shift of the resonance, the density effect, the broadening of the line, the artifact due to the ion blockade are reproduced with the simulations we describe. For varying initial atomic densities the amplitude of the energy shift due to the nearest neighbor is monitored which confirms its dominant role in the dipole shift. We could also use the Kinetic Monte Carlo model to analyse other experiments than ours as in figure \ref{fig:longexc} A) \cite{Singer}. As another simple example, the number of atoms excited to a Rydberg state in our simulation is analyzed using the the Mandel $Q$ parameter and the statistics follow a sub-Poissonian distribution as described in \cite{RaithelPRL05, RaithelPRL05erratum, Robicheaux2005, Ates2006}. The density matrix model can be a good tool to model recent experiments of coherent excitation of Rydberg atoms \cite{Pfau1,Pfau2,Pfau3,reetzlamour-2007}.
The next step will be the modelling of the F\"{o}rster case \cite{VogtPRL2006} where the coherences between pairs of atoms are expected to play a dominant role.

\ack{This work is in the frame of "Institut francilien de
recherche sur les atomes froids" (IFRAF). This theoretical development could aslo be applied to the CORYMOL experiment supported by an ANR grant (No. NT05-2 41884). The authors
thank Thomas F. Gallagher and Jianming Zhao.}

\appendix
\section{KMC method for solving rate equations}
\label{appendixA}
   
\subsection{The master,  kinetic or (reaction-)rate equation}

There are innumerable instances, in physics and in other sciences, where a system evolves in time through many competing
internal stochastic processes. For instance,
many classical and quantum physical problems can be reduced to the form of a master equation: 
\begin{equation}
    \frac{dP_k}{dt} =  \sum_{l=1}^N \Gamma_{k l} P_l - \sum_{l=1}^N  \Gamma_{l k}  P_k 
    \label{master Eq.}
\end{equation}
  This master equation, sometimes called  kinetic or (reaction-)rate equation is a phenomenological set of coupled first-order differential equations describing the time evolution of the probability $P_k$ of a system to occupy each one of a discrete set of states numbered by $k$.  In probability theory, this identifies the evolution as a continuous-time Markov process. Each process occurs at a certain average rate $\Gamma_{l k}$, which may either
be constant in time, or dependent on how the system has evolved up to that time. The goal of this appendix is to describe why Kinetic
Monte Carlo methods are a standard means of modelling such  problems, especially
when one wishes to model the evolution of the system over periods of time much longer
than those accessible by direct simulation. 

\subsection{Solving the master equation}

Following Gillespie \cite{Gill77} we can distinguish, among several competing methods commonly used to solve the master equation, two major approaches: The \textit{deterministic approach} which regards the time evolution as a continuous, wholly predictable process governed by a set of coupled, ordinary differential equations (the "reaction-rate equations") and  the \textit{stochastic
approach} which regards the time evolution as a kind of random-walk process which is governed by a single differential equation (the "master equation") governing
the time-dependent behavior rather than a fixed probability distribution.

\subsubsection{Deterministic approach}
~\\
The deterministic approach is based on the fact that equation (\ref{master Eq.})
can be written as a matrix ordinary differential equations 
$  \frac{d \mathbf{P}  }{dt} = \mathbf{\Gamma} \mathbf{P}$ 
The formal solution 
\begin{equation}
\mathbf{P} (t) = \mathbf{P} (0) \exp\left( \int_0^t  \mathbf{\Gamma} (t') d t' \right) \label{integr_ME}
\end{equation}
can be obtained for instance by using Direct diagonalization algorithms \cite{2003JChPh.11912741F}.
A second deterministic approach to the time-dependent population
distribution comes from the integrand form of the master equation:
\begin{equation}
\mathbf{P} (t) = \mathbf{P} (0) + \int_0^t  \mathbf{\Gamma} (t') \mathbf{P} (t') d t'
\end{equation}
Explicit
numerical integration can in principle be
achieved by different numerical integration schemes.
All these methods  require the explicit
formation of the matrix and the computational effort is dominated by $N^3$ terms \cite{2003JChPh.11912741F}.

The deterministic approach
is simply the exact time-evolution
for the function $\mathbf{P}$.
However, the stochastic probabilistic formulation has often a stronger physical basis, especially in the quantum world or for non equilibrium systems, than the deterministic formulation. 
  Instead of the deterministic approach which deals only with one possible ''reality'' of how the process might evolve under time, in a stochastic or random process there is some indeterminacy in its future evolution described by probability distributions. This means that even if the initial condition (or starting point) is known, there are many possibilities where the process might go to, but some paths are more probable and others less.
This  approach correctly accounts for the inherent fluctuations
and correlations that are necessarily ignored in the deterministic formulation. 
In addition, as we shall see, this point of view, opens the way to stochastic simulation algorithms, such as the Kinetic Monte Carlo one, making exact numerical calculations which are much faster $O(N)$ than the deterministic reaction-rate algorithms.

\subsubsection{Monte Carlo algorithms}
~\\
Monte Carlo refers to a broad class of algorithms that solve problems through the use of random numbers   \cite{10.1109/MCSE.2006.34, Gardiner2004}. They first emerged in the late 1940's and 1950's as electronic computers came into use. The most famous of the Monte Carlo methods is the Metropolis algorithm (sometimes called Monte Carlo Markov chain methods), offering an elegant and powerful way to generate equilibrium properties of physical systems \cite{Siepmann}.
For many years, researchers thought Monte Carlo methods could not be applied to molecular dynamics simulations because it seems necessary  to follow individual motion and/or interactions \cite{NIC1,NIC2,NIC3}.  However some Monte Carlo algorithms of this type exist; they are sometimes called Coarse-Grained methods \cite{Attard}.
The simplest algorithm of this kind, sometimes called fixed time step algorithm \cite{2003cond.mat..3028J},  is based on the first-order formula $\mathbf{P} (t+d t ) = \mathbf{P} (t) +  \mathbf{\Gamma} (t) \mathbf{P} (t) d t $ i.e.
\[  P_k(t+dt) = P_k(t) - \sum_{l=1}^N  \Gamma_{l k}(t) P_k(t) dt   +  \sum_{l=1}^N \Gamma_{k l}(t)  P_l (t) dt \]
A similar scheme is, for instance, used in stochastic quantum simulation in the 
	Quantum Monte Carlo wave-function approach \cite{citeulike:913853} because the Lindblad equation in quantum mechanics is a generalization of the master equation describing the time evolution of a density matrix.

To illustrate the fixed time step method \cite{2003cond.mat..3028J}, let's assume that at time $t$ the system is in state $k$: $P_k=1$ and $P_l=0$ for $l\neq k$. The algorithm 
 consists of choosing a small $d t$,  and for each possible reaction $k\rightarrow l$ generating a random number $r$ between $0$ and $1$. If $r<\Gamma_{l k}(t)  dt$ the system changes configuration and evolves to state $l$ at time $t+ d t$ (a quantum jump occurs in the stochastic quantum simulation terminology). The main disadvantage is that  $d t$ has to be small enough to maintain accuracy and such that at most one reaction occurs during each time step: meaning  $\Gamma_{l k}(t)  dt \ll 1$. Several steps are then needed before effectively doing the evolution. As we shall see, especially for time independent $\Gamma_{l k}$ rates, this Monte Carlo algorithm is very inefficient compared to the kinetic Monte Carlo algorithm which ensures at each time step an evolution of the system.

\subsection{The Kinetic Monte Carlo (KMC) method}

\subsubsection{Derivation}
~\\
 The Kinetic Monte Carlo algorithm is also known as the residence-time algorithm, the n-fold way, the Bortz-Kalos-Liebowitz (or BKL) algorithm \cite{1975JCoPh..17...10B}, the dynamical Monte Carlo method \cite{1991JChPh..95.1090F}, the Gillespie algorithm \cite{1976JCoPh..22..403G,Gill77}, the Variable Step Size Methods (VSSM)
\cite{2003cond.mat..3028J} (by comparison with the fixed time step method) ..., depending on the physical or chemical context. For other reviews of the KMC algorithm see \cite{2003cond.mat..3028J,Gardiner2004,Siepmann,Chatterjee}. Some minor subtle changes between these algorithms exist, but we will describe here only 3 types of KMC algorithms, namely the Kinetic Monte Carlo (KMC) algorithm, the First Reaction Method (FRM) and the Random Selection Method (RSM).
    
The Kinetic Monte Carlo method is a Monte Carlo method intended to simulate the time evolution of independent (non correlated) Poisson processes \cite{2003cond.mat..3028J,Voter}. This means that the KMC method solves the master equation and is therefore of great interest, for instance for relaxational
processes and transport processes on mesoscopic to macroscopic
time scales. Indeed, the KMC algorithms
are able to model the evolution of the system over periods of time
very much longer than those accessible by direct simulation such as molecular dynamics. Surprisingly enough, up to now it has been
more or less limited to the study of
chemical reactions, surface or cluster physics (diffusion, mobility, vacancy motion, transport process, epitaxial growth, dislocation, coarsening,  ...).

Due to the Markovian behavior the system
 loses its memory of how it entered state $k$ at time $t$.
Therefore, in order to simulate the
stochastic time evolution of such a reacting system, i.e. in order to move
the system initially at time $t$ in state $k$ forward in time, we just need to know when  the next reaction will occur, and what
kind of reaction it will be \cite{Gill77}. We then need to determine the probability distribution function $p(t)$ for the time of the
first system change.
From equation (\ref{integr_ME})
the probability that the system
has not yet escaped from state $k$ at time $t'$ is given by
$p_{\mathrm{survival}} (t') = \exp \left( \int_t^{t'} \Gamma_{k}(\tau) d \tau  \right) $, where $\Gamma_k = \sum_l \Gamma_{l k}$ is the total  rate from state $k$. Thus $1- p_{\mathrm{survival}} (t')$
gives the probability that the
system has been modify at time $t'$, which is exactly the integral $\int_t^{t'} p(\tau) d \tau$. The first-passage-time distribution in then found by differentiation:
\begin{equation}
p(t') = \Gamma_k(t') \exp \left(  -\int_t^{t'} \Gamma_k(\tau) d \tau  \right)
\label{proba_distri}
\end{equation}
which is characteristic of the Poissonian nature of the process and is the starting point of the KMC algorithm.
We then know when the next reaction will occur. We just have to find what
kind of reaction it will be.
At time $t'$ a reaction takes place, so just before $t'$ the system is still in state $k$ and according to the master equation (\ref{master Eq.}) the probability that the system will be in configuration
$l$ at time $t' + dt'$ is  $\Gamma_{lk}(t') d t'$, where $dt'$ is a small time interval. 
We therefore have to
generate a new configuration $l$ by picking it out of all possible new configurations with a probability proportional to
$\Gamma_{lk} (t')$. 

The algorithm is based on our ability to
 generate a time $t'$ from equation (\ref{proba_distri}) when the first reaction actually occurs and on our ability to choose the correct reaction $l$.
 
 The $t'$ choice 
 can be done  by the inverse transform method  \cite{PDBook} which is based on the fact that
 if $t'$ is chosen with probability density $p(t')$, then the integrated probability
up to point $t'$, $\int_t^{t'} p(\tau) d \tau = 1- p_{\mathrm{survival}} (t')$, is itself a random variable which will occur with uniform probability
density on $[0, 1]$. In conclusion,
to
 generate a time $t'$ with probability density $p(t')$ we just have to solve for $t'$ the equation
$r =  \exp \left(  -\int_t^{t'} \Gamma_k(\tau) d \tau  \right)$
where $r$ is a unit interval uniform random number. This is exact and totally trivial when the rates are time independant explaining why that KMC method is so powerful in this case. The following choice of the reaction $l$ is done by randomly choosing $r'\in]0,1]$ and by finding the $l$ for which $R_{l - 1} < r' R_N \leq  R_l$ where
 $R_j = \sum_{i=1,j} \Gamma_{i k} (t')$. At this time $t'$ we are in the same situation as when we started the simulation,
and we can proceed by repeating the previous steps. 
The resulting algorithm is not reproduced in this Appendix because it is described in section \ref{KMC_algo}.

\subsubsection{Algorithms}
~\\
This KMC algorithm  is clearly $O(N)$, since at least the third step has a sum over $N$ elements.
It is beyond the scope of this annexe to discuss all alternative methods but it is sometimes possible to improve the speed of the algorithm, for instance if several rates are similar, using a binning or weighted methods to reduce for instance to $O(\log N)$ the complexity \cite{1988SeScT...3..594M,1995PhRvE..51..867B,Ovcharenko,2003cond.mat..3028J}.

It is of interest to mention at least the First Reaction Method (FRM) and the Random Selection Method (RSM) \cite{2003cond.mat..3028J}. Depending of the type of reaction in the system these algorithms can be useful.
The FRM, also called Discrete Event Simulation in computer science, consists of choosing the first occurring reaction, meaning choosing the smallest time $t'_l$, and the corresponding reaction number $l$, from the formula $\int_{t}^{t'_l} \Gamma_{l k}(\tau) d \tau = -\ln r_l$  where the $r_l \in ]0, 1]$ are $N$ independent random numbers. As the KMC one this algorithm generates an exact evolution of the system but
  is usually less efficient  because
it necessitates a random number for each possible reaction, whereas the KMC advances the system to the next state with just two
random numbers. 

The RSM can be used only when the rates $\Gamma_{l k}(t) = \Gamma_{l k}$ are time independent, as  for Poisson processes (such as radiative lifetime, radiative decay rate, ...).  The RSM 
consists of evolving  the system up to the   time $t' = t -(\ln r)/\Gamma_{\mathrm{max}}  $  where $r \in]0, 1]$ is a random number and  $\Gamma_{\mathrm{max}} = \max_l \Gamma_{l k}$ is the maximal possible rate. Then choosing randomly a possible reaction $l\in[1,N]$ and accepting the reaction with
probability $\Gamma_{l k}/\Gamma_{\mathrm{max}}$.
If the reaction is accepted, the configuration is changed. Contrary to the FRM or the KMC algorithms it does not necessarily imply a system evolution at each time step. But, here again 
this algorithm generates an exact evolution of the system. The RSM
 is optimized for system having  just one (or a small number of) type of reaction because it is then
of $O(1)$ complexity !

Following \cite{2003cond.mat..3028J}, KMC is generally the best method to use unless the number of
reaction types is very large. In that case use FRM. If you have a type of reaction that
occurs almost everywhere, RSM should be considered. Simply doing the simulation
with different methods and comparing is of course the best.

\subsection{Conclusion}
The KMC algorithm is a stochastic algorithm generating quasi classical trajectories, "i.e."
creating a Markov chain representing the exact evolution of the system in the sense that it will be statistically indistinguishable from
an exact dynamics simulation. 
 Indeed, each system configuration $l$ is reached with its real physical probability. Unlike most procedures such as the often used fixed time step method
for numerically solving the deterministic reaction-rate equations, this algorithm never approximates infinitesimal
time increments $dt$ by finite time steps $\Delta t=t'-t$.

The fact that the mechanisms and so the rates have to be known in advance is the main limitation of the use of the KMC method.
If the rate have to be modified in an unpredictable fashion during the free time evolution of the system, i.e. between two reactions, the fixed time step method has to be preferred. 
 However for several physical system this is not the case and the KMC or FRM algorithms can be used. As the interaction between particles of a system depends often of the distance between particles, the KMC methods can be advantageously used when the motion of the particles is slow. Ultimately when the rates are time independent (which does not mean that they are constant, because they often have to be recalculated after each system evolution) KMC and RSM approaches are very powerful.

\section{Simple method to solve the N-body problem}
\label{appendixB}

\subsection{Introduction}
A large number of physical systems can be studied by simulating the interactions between the particles constituting the system. In a typical system each particle influences every other particle. The interaction is often based on an inverse square law such as Newton's law of gravitation or Coulomb's law of electrostatic interaction but, as in our case, a more complex anisotropic interaction with an inverse higher power law dependence might exist. Examples of such physical systems can be found in astrophysics, plasma physics, molecular dynamics and fluid dynamics. Since the simulation involves following the trajectories of motion of a collection of N particles, the problem is termed the N-body problem. 

Since it is not possible to solve the equations of motion for a collection of many particles in closed form, iterative methods are used to solve the N-body problem. At each discrete time interval, the force on each particle is computed and this information is used to update the position and velocity of each particle. A straightforward computation of the forces requires $O(N^2)$ work per iteration. The rapid growth with $N$ limits the number of particles that can be simulated by this method.  Several approaches, especially that by Aarseth \cite{Aarseth}, have been used to reduce the complexity per iteration and to speed up the calculation, for instance each particle is followed with its own integration step.
Non full N-body codes also exist transforming the problem imposing for instance a grid on the system of particles and computing cell-cell interactions. These are known as  hierarchical methods, or tree methods such as the Barnes-Hut one where a Ahmad-Cohen neighbor type of scheme is used which updates less frequently the non neighbor force than the neighboring one. Finally, multipole expansion methods have also been developed as well as Monte Carlo algorithms for very large number of particles using a set of representative ''macro'' particles (not point-masses) like in Fokker-Planck or gaseous methods with smooth potentials after the pioneering work of H\'enon \cite{Henon}.
 The required CPU time scales with the number $N$ of particles as $N \ln(N)$ (for a given number of relaxation times), while the scaling is $N^k$ with $k$ of order $2-3$ for direct N-body codes. 
It is beyond the scope of our article to discuss all the methods:
Particle-Particle (PP),
Particle-Mesh (PM),
Particle-Particle/Particle-Mesh (P3M),
Particle Multiple-Mesh (PM2),
Nested Grid Particle-Mesh (NGPM),
Tree-Code (TC) Top Down or Bottom Up,
Fast-Multipole-Method (FMM),
Tree-Code Particle Mesh (TPM),
Self-Consistent Field (SCF),
Symplectic Method, ...
 Very good references, discussing also their stability or their complexity can be found on the web site http://www.manybody.org/ or in the references \cite{2005astro.ph..3600H,HeggieHut,Aarseth}. 

\subsection{Choice of an N-body integrator}
We have based our code on  
a series of books centered around N-body simulations "The Art of Computational Science" by Piet Hut and Jun Makino http://www.artcompsci.org/. Because of the small number of particles involved we have used a very simple algorithm. Another reason is that the KMC algorithm is the one which limits the cpu time. Finally the forces we use (see equation (\ref{eq:dipdip})) are not accurate for all interparticles distances. Therefore, it is not necessary to have a very powerful N-body code.
In order to deal with a very  simple and versatile code, the code is written in C++ in a completely stand-alone fashion based on the N-body
 ''Starter Code for N-body Simulations''. \footnote{http://www.ids.ias.edu/\~piet/act/comp/algorithms/starter/index.html. In order to use the 3D vector formulation we have based our code on the CERN library CLHEP (A Class Library for High Energy Physics).} 

The key part of our code is the N-body integrator.
Hamiltonian systems are not structurally stable against non-Hamiltonian perturbations \cite{Li2007,Leimkuhler}. The ordinary numerical approximation to a Hamiltonian system obtained from an ordinary numerical method does introduce dissipation, with completely different long-term behavior, since dissipative systems have attractors and Hamiltonian systems do not. This problem has led to the introduction of methods of symplectic integration for Hamiltonian systems, which do preserve the features of the Hamiltonian structure by arranging that each step of the integration be a  canonical or  symplectic transformation. Many different symplectic algorithms have been developed and discussed \cite{Cartwright}.  
Symplectic integrators tend to have much better total energy
conservation in the long run.
 Finally, to save computational cost, most often
one must adopt a quite large $\Delta t$ step and
higher-order (local truncation error) algorithm. However, because of the computational round-off error and due to their smaller stability domain than the lower-order algorithm at practical $\Delta t$, high-order algorithms pushes the machine precision limit \cite{Li2007} and algorithms are generally not good to go beyond 3rd or 4th order. Finally a high-order predictor-corrector integrators  have usually a better performance than the
symplectic integrators at large integration timestep.

For all these reasons one very popular N-body integrator is the fourth (local) order ``Hermite'' predictor-corrector scheme by Makino and Aarseth \cite{HeggieHut,Aarseth,2007arXiv0711.0643P}.

However the Hermite algorithm requires knowledge of the time derivatives of the acceleration (sometimes known as jerk), which can be difficult to evaluate. For that purpose we use
a simple but efficient algorithm, the so called leapfrog-Verlet-St\'ormer-Delambre algorithm used in 1791 by Delambre, and rediscovered many times, 
and recently by the French physicist Loup Verlet in 1960s for molecular dynamics. 
The position Verlet method does not store explicit velocities, allowing it to be extremely stable in cases where there are large numbers of mutually interacting particles. It is, of $O((\Delta t)^4)$ for local truncation error for position and $O((\Delta t)^2)$ for velocity. We are interested in having accuracy in position and velocity so we use the so called
Velocity Verlet method which is of  $O((\Delta t)^3)$ accuracy for both position and velocity for a $\Delta t$ timestep. 
This leapfrog integrator often turns out to be more accurate than expected 
from a simple second-order integrator \cite{1986PaAcc..19...37H}. 
The `unreasonable' accuracy stems from its symmetry properties under time invariance due to its simplectic structure.  
The scheme of the algorithm is the following.

\begin{eqnarray}
\mathbf{r}_i (t+ \Delta t) &=& \mathbf{r}_i (t) + \mathbf{v}_i (t) \Delta t + \frac{1}{2}
\mathbf{a}_i(t) (\Delta t)^2 \\
\mathbf{v}_i (t+ \Delta t) &=& \mathbf{v}_i (t) + \frac{1}{2} \left(
\mathbf{a}_i(t) + \mathbf{a}_i(t+\Delta t)  \right)(\Delta t)
\end{eqnarray}
This has also the big advantage that accuracy can be improved by using  higher order symplectic integrators such as the one by Yoshida
\cite{1990PhLA..150..262Y}.
\\
\bibliographystyle{unsrt}
\bibliography{KMCbiblio_bib}
\end{document}